\documentclass[nofootinbib,aps,a4paper,10pt,letterpaper,superscriptaddress,
twocolumn,showkeys,times,eqsecnum]{revtex4-1}
\usepackage{amsmath}
\usepackage{amsmath,tabstackengine}
\usepackage{amsfonts}
\usepackage{times}
\usepackage{tabularx}
\usepackage{framed}
\usepackage{nccmath} 
\usepackage[utf8]{inputenc}
\usepackage{amsmath}
\usepackage{colortbl}
\usepackage{array}       
\usepackage{tabularx}     
\usepackage{lipsum} 
\usepackage{mathtools} 
\usepackage{array}
\usepackage{multirow}
\usepackage{makecell}
\usepackage{booktabs}
\usepackage{graphicx}
\usepackage{multirow}
\usepackage{color}
\usepackage{braket}
\usepackage{dcolumn}
\usepackage{siunitx}
\usepackage{bm,url}
\usepackage{colortbl}
\usepackage[linktocpage]{hyperref}
\usepackage{subfigure}
\usepackage{amsfonts}
\usepackage[left=2cm,right=2cm,top=2.5cm,bottom=2.5cm]{geometry}
\usepackage[usenames,dvipsnames,svgnames]{xcolor}  
\usepackage{hyperref}    
\definecolor{oxfordblue}{rgb}{0.0, 0.13, 0.28}
\definecolor{burgundy}{rgb}{0.5, 0.0, 0.13}
\definecolor{darkolivegreen}{rgb}{0.33, 0.42, 0.18}
\definecolor{darkblue}{rgb}{0,0,0.5}
\definecolor{richcarmine}{rgb}{0.84, 0.0, 0.25}
\definecolor{darkblue}{rgb}{0,0,0.5}
\definecolor{bluer}{rgb}{0.00,0.50,0.75}{}
\hypersetup{colorlinks=true, citecolor=red, linkcolor=blue,
	urlcolor = magenta, filecolor=magenta}

\newcommand\be{\begin{equation}}
\newcommand\ee{\end{equation}}
\newcommand\bea{\begin{eqnarray}}
\newcommand\eea{\end{eqnarray}}
\newcommand\bseq{\begin{subequations}} 
\newcommand\eseq{\end{subequations}}
\newcommand\bcas{\begin{cases}}
\newcommand\ecas{\end{cases}}

\begin{document}
	
\title{	Atomic clocks and gravitational waves as probes of non-metricity}
 
	\author{Mohsen Khodadi}
	\email{khodadi@kntu.ac.ir}
\affiliation{School of Physics, Institute for Research in Fundamental Sciences (IPM),	P. O. Box 19395-5531, Tehran, Iran}
	\affiliation{School of Physics, Damghan University, Damghan 3671641167, 
Iran}
	\affiliation{Center for Theoretical Physics, Khazar University, 41 Mehseti 
Str., AZ1096 Baku, Azerbaijan}

 \author{Emmanuel N. Saridakis}
\email{msaridak@noa.gr}
 \affiliation{Institute for Astronomy, Astrophysics, Space Applications and 
Remote Sensing, National Observatory of Athens, 15236 Penteli, Greece}
\affiliation{Departamento de Matem\'{a}ticas, Universidad Cat\'{o}lica del 
	Norte, Avda. Angamos 0610, Casilla 1280, Antofagasta, Chile}
\affiliation{CAS Key Laboratory for Research in Galaxies and Cosmology, 
	School 
	of Astronomy and Space Science,
	University of Science and Technology of China, Hefei 230026, China}

\begin{abstract}
Non-metricity provides a natural extension of Riemannian geometry, yet its
experimental signatures remain largely unexplored.
In this work we investigate how spacetime non-metricity can be probed through
high-precision observations, focusing on atomic clocks and gravitational waves
as complementary tools.
Working within Weyl geometry as a minimal realization of vectorial
non-metricity, we formulate observable effects in a gauge-invariant manner and
show that they are associated with path-dependent length transport governed by
the Weyl field strength. 
We derive constraints from atomic-clock experiments and demonstrate that,
although gravitational waves do not directly source the Weyl field at linear
order, its dynamical contribution induces a backreaction on gravitational-wave
propagation, leading to an anomalous strain.
As a result, the absence of deviations from General Relativity in current
gravitational-wave observations already places meaningful and strong 
constraints on
dynamical non-metric degrees of freedom, within the phenomenological
classical framework considered.

\vspace{0.5cm}
\end{abstract}

\maketitle

\section{Introduction}

General Relativity (GR) describes gravity as spacetime curvature encoded in a metric
$g_{\mu\nu}$ and a torsion-free, metric-compatible Levi-Civita connection.
Metric compatibility, $\nabla_\lambda g_{\mu\nu}=0$, ensures that lengths and
angles are preserved under parallel transport and is built into the standard
geometric formulation of Einstein gravity.
However, this condition is not imposed by fundamental principles.
Allowing a more general affine connection gives rise to non-metricity,
$\nabla_\lambda g_{\mu\nu}\neq 0$, which plays a central role in metric-affine
extensions of gravity 
\cite{Hehl:1994ue,Blagojevic:2002du,Puetzfeld:2007hr,Iosifidis:2019dua}.

A particularly simple and well-controlled realization of non-metricity is
provided by Weyl geometry, in which the covariant derivative of the metric is
proportional to a vector field, namely
$\tilde{\nabla}_\lambda g_{\mu\nu}=-\alpha\,\omega_\lambda g_{\mu\nu}$,
with $\omega_\mu$ the Weyl gauge field and $\alpha$ a coupling constant
\cite{Weyl:1918ib,Scholz:2011za}.  

At this point, it is useful to recall briefly the concept of 
hypermomentum, which is a central quantity in metric-affine theories of gravity 
\cite{Hehl:1994ue,Blagojevic:2002du}. The hypermomentum tensor is defined as
\begin{equation}
\Delta^{\mu\nu}{}_\lambda \equiv
-\frac{2}{\sqrt{-g}}\frac{\delta S_M}{\delta \Gamma^\lambda_{\mu\nu}},
\end{equation}
and characterizes the coupling of matter to the independent affine connection. 
It naturally decomposes into spin, corresponding to the antisymmetric part, 
shear, corresponding to the traceless symmetric part, and dilation, 
corresponding to the trace part, which source torsion and non-metricity, 
respectively \cite{Puetzfeld:2007hr,Iosifidis:2019dua}. In the Weyl, or 
vectorial non-metricity, case, matter sources the Weyl field through the 
dilation current, corresponding to the matter current $J_{\mathrm{matter}}^\nu$ 
in the Proca equation \cite{Weyl:1918ib}. Beyond the central role of hypermomentum tensor in extended gravities and metric-affine theories \cite{Obukhov:1987yu,GRG1993,Obukhov:1996ka,Obukhov:2023gyr}, it finds important applications in the relativistic fluid dynamics 
\cite{Obukhov:1993pt,Obukhov:1996mg,Iosifidis:2023kyf,Obukhov:2023yti}, and cosmology \cite{Obukhov:1997zd,Iosifidis:2020gth,Iosifidis:2021xdx,Iosifidis:2022xvp,Iosifidis:2024bsq}. The connections of hypermomentum with field theory studied in \cite{Ariki:2018hnz} (see also \cite{Iosifidis:2025sjx}), too.

In modern formulations, Weyl-invariant and quadratic gravity theories compose a
consistent framework in which the Weyl field becomes dynamical and typically
acquires a mass through a Stueckelberg-like mechanism
\cite{Ghilencea:2018dqd,Ghilencea:2019jux,Ghilencea:2022lcl}.
Despite their theoretical appeal and detailed
cosmological and black hole
applications 
\cite{Sola:1988nz,Mannheim:1990ya,La:1991nk,Elizondo:1994vh,Bronnikov:1997gj,
Edery:1997hu,Klemm:1998kf,Boulanger:2001he,Pireaux:2004xb,Pireaux:2004id,
Flanagan:2006ra,Edery:2006hg,Lobo:2008zu,Sultana:2010zz,Percacci:2011uf,
Dengiz:2011ig,Tanhayi:2012nn,Sultana:2012qp,Deruelle:2012xv,Said:2012xt,
Cattani:2013dla, 
Wheeler:2013ora,Quiros:2014hua,Myung:2014jha,Cusin:2015rex,Mureika:2016efo, 
Oda:2016psn, Zinhailo:2018ska,
Ghilencea:2018thl, Gomes:2018sbf,
Ghilencea:2020rxc,Takizawa:2020dja,Jawad:2020wlg,Xu:2020yeg,Geiller:2021vpg,
Yang:2022icz, Harko:2022iva,
Hell:2023rbf,
Karananas:2021gco,Roumelioti:2024lvn,Gialamas:2024iyu,Lima:2024cys,
Momennia:2025ngm,Khodadi:2025gtq,Lima:2025ztp,Kusano:2025rly,Gomes:2025oxz,
Bahamonde:2025xum}, 
such theories face a 
long-standing
phenomenological difficulty, namely that  non-metricity effects are 
very hard to be
detected 
\cite{Lobo:2018zrz,Delhom:2020vpe,Iosifidis:2023eom,Obukhov:2024evf,
Iosifidis:2025ano}.

We mention here that the existence of non-metricity does not 
necessarily imply a violation of local Lorentz invariance. In metric-affine 
frameworks the non-metricity tensor $Q_{\lambda\mu\nu} = 
\nabla_\lambda g_{\mu\nu}$ describes properties of the affine connection and the 
transport of lengths and angles, while local Lorentz symmetry refers to the 
invariance of the local tangent-space structure of the theory. These are 
logically distinct conditions: the light cone structure, determined by 
$g_{\mu\nu} ds^\mu ds^\nu = 0$, remains unchanged by 
non-metricity. However, the distinction is more subtle in effective 
descriptions. As Obukhov and Hehl~\cite{Obukhov:2024evf} have shown, certain 
realizations of non-metricity can generate operators in the matter sector that 
effectively break Lorentz invariance, providing a geometric origin for such 
effects. In the present work, we focus on a different aspect: we consider a 
geometric realization of vectorial non-metricity within Weyl geometry and 
investigate its possible observational signatures without introducing explicit 
Lorentz-violating operators in the matter sector. Our analysis therefore 
addresses the phenomenology of geometric non-metricity itself, rather than 
Lorentz-violation scenarios, while acknowledging the potential connections that 
exist in more general frameworks.
Consequently, the extremely strong experimental bounds on Lorentz and CPT violation (see e.g. the Particle Data Group review~\cite{ParticleDataGroup:2024cfk}) do not directly exclude the possibility of non-metricity in a geometric framework, although they may constrain certain effective descriptions derived from it.

Weyl’s original attempt to unify gravity 
and 
electromagnetism through a geometrical scale connection was historically 
criticized, most notably by Einstein, due to the so-called ``second clock 
effect'', which would lead to path-dependent clock rates. This issue has been 
widely discussed in the literature (see e.g. the historical review in 
\cite{Penrose:2017wxd}). In modern gravitational studies, however, 
Weyl geometry is typically employed in a different spirit: the Weyl one-form 
$\omega_\mu$ is treated as an independent geometric degree of freedom 
representing vectorial non-metricity, rather than being identified with the 
electromagnetic field. In this context Weyl geometry provides a minimal and 
well-defined framework for investigating the phenomenology of vectorial 
non-metricity.

Now, in Weyl geometry, non-metricity may manifest itself as a path dependence 
of
lengths under parallel transport, historically associated with the ``second
clock effect'' \cite{Scholz:2011za}.
The absence of such an effect in laboratory experiments has often been
interpreted as placing severe constraints on long-range Weyl fields or as
suggesting that non-metric degrees of freedom are irrelevant at accessible
scales.
This interpretation, however, implicitly assumes that non-metricity must reveal
itself through local and quasi-static measurements.
In this work we argue that this viewpoint is overly restrictive and that
non-metricity can be probed efficiently by exploiting recent advances in
precision metrology and gravitational-wave (GW) astronomy.

Two developments make this possibility particularly timely.
Firstly, atomic clocks have reached fractional frequency uncertainties at the
$10^{-18}$ level and below, enabling precision tests of gravitational redshift,
local position invariance, and possible variations of fundamental constants
\cite{Ludlow:2015gnc,Uzan:2010pm,Safronova:2017xyt}.
Secondly, gravitational-wave detectors measure the strain $h=\Delta L/L$ with
extraordinary sensitivity, providing direct access to the dynamical sector of
gravity.
Since the first detection GW150914 
\cite{LIGOScientific:2014pky,LIGOScientific:2016aoc}, and subsequent
events such as GW170817 and GW190521
\cite{LIGOScientific:2017vwq,LIGOScientific:2020iuh}, GW
astronomy has enabled precision tests of gravity in regimes that were previously
inaccessible, with further dramatic improvements expected from next-generation
and space-based observatories
\cite{Sathyaprakash:2012jk,Evans:2021gyd,Kawamura:2020pcg}. In this context, 
the 
absence of observed deviations from GR in current 
gravitational-wave signals should not be interpreted as a lack of sensitivity 
to 
non-metricity, but rather as a meaningful constraint on the dynamical 
contribution of non-metric degrees of freedom to GW propagation.

It is important to clarify the scope of the present work. Our goal is not to 
develop or revive the full framework of metric-affine gravity (MAG), but 
rather to investigate the observational consequences of a specific and minimal 
realization of vectorial non-metricity. In particular, we restrict our analysis 
to the Weyl-type sector in which the non-metricity tensor takes the form 
$Q_{\lambda\mu\nu}=2\,\omega_\lambda g_{\mu\nu}$, characterized by a single 
vector field $\omega_\mu$. This sector provides a well-defined geometrical 
framework that captures the essential features of vectorial non-metricity, 
while remaining sufficiently simple to allow phenomenological analysis.
Within this setup our aim is to explore whether such a geometric degree of 
freedom could leave observable signatures in precision experiments, such as 
atomic-clock measurements and gravitational-wave observations. Thus, the 
approach 
  remains purely phenomenological and focuses on deriving observational 
constraints on vectorial non-metricity rather than constructing a general 
metric-affine theory.

The purpose of the present work is to demonstrate that atomic clocks and
gravitational waves provide robust and complementary probes of vectorial
non-metricity.
Our approach is twofold.
On the one hand, we formulate the observable consequences of non-metricity in a
manifestly gauge-invariant manner, expressing measurable effects in terms of the
Weyl field strength rather than the gauge field itself.
On the other hand, we investigate how gravitational waves can probe 
non-metricity
through the dynamical response of the Weyl sector within Weyl quadratic gravity.
We show that, although direct sourcing of the Weyl field by GWs
is absent at leading order, a consistent dynamical treatment, including the
energy-momentum tensor of the Weyl field, opens a non-trivial channel through
which non-metricity can affect gravitational-wave propagation and the measured
strain.

At this point it is important to clarify the scope of our analysis.
In general metric-affine geometry, the non-metricity tensor
$Q_{\lambda\mu\nu}\equiv\nabla_\lambda g_{\mu\nu}$ contains several irreducible
components, and its physical implications depend sensitively on both its
tensorial structure and the way matter couples to the connection 
\cite{CANTATA:2021asi}.
Weyl geometry corresponds to the special case of purely vectorial non-metricity,
leading to a universal rescaling of lengths under parallel transport and to
particularly clean, gauge-invariant observables.
By contrast, in symmetric teleparallel and $f(Q)$ theories non-metricity is
generically non-vectorial and gravity is encoded in the scalar $Q$
\cite{Nester:1998mp,BeltranJimenez:2018vdo}.
While the conceptual strategy developed in the present work is more general, 
the concrete
realization and constraints derived here are specific to the vectorial 
non-metricity sector of Weyl geometry.

We note that alternative proposals for probing non-metricity have been explored in the literature. In particular, authors in Ref.~\cite{Foster:2016} have derived constraints on non-metricity by exploiting high-precision searches for Lorentz violation. Working within the Standard-Model Extension (SME) framework, they obtained bounds on the 40 independent components of the non-metricity tensor down to levels of order $10^{-43}$ GeV, using observations ranging from atomic clocks to cosmological birefringence. This approach is complementary to ours: while Foster et al.~\cite{Foster:2016} constrain non-metricity through its effective Lorentz-violating couplings to fermions and photons in the SME, we focus on the geometric phenomenology of vectorial non-metricity in Weyl geometry, deriving constraints from atomic clocks and gravitational waves without invoking explicit Lorentz violation. The two approaches thus probe different aspects of non-metricity and together provide a more complete picture of its possible observational signatures.

The structure of the paper is as follows.
In Section~\ref{Mathematicaldescription} we derive the path dependence of 
physical
lengths in Weyl geometry and identify the relevant observables.
In Section~\ref{atom} we discuss atomic clocks as probes of non-metricity.
Section~\ref{GWprobes} is devoted to gravitational waves as probes of non-metricity, where we examine the theoretical challenges in coupling gravitational waves to the Weyl field, identify an effective long-wavelength source term, and derive exploratory bounds under a specific truncation of the field equations. We conclude in Section~\ref{Conclusions}.

 \section{Path-dependent length transport in Weyl geometry}
 \label{Mathematicaldescription}

In this section we use the mathematical description of ``frame dragging of 
length'' in order to
derive constraints on the non-metricity scale.
We consider vectorial non-metricity within Weyl geometry, defined through 
\cite{Weyl:1918pdp,Romero:2012hs,Wheeler:2018rjb,Kouniatalis:2024gnr}
\be\label{met}
\tilde{\nabla}_{\mu}g_{\alpha\beta} = -\alpha\,\omega_{\mu}g_{\alpha\beta},
\ee
where $\tilde{\nabla}_{\mu}$ denotes the covariant derivative associated with 
the Weyl connection
$\tilde{\Gamma}$, $\alpha$ is the Weyl gauge coupling, and $\omega_\mu$ is the 
Weyl gauge field. We emphasize that the coupling constant \(\alpha\) is a free parameter of the geometric theory; in phenomenological applications it may be absorbed into the definition of \(\omega_{\mu}\), but we keep it explicit for clarity.

Within this framework, the norm of a vector $u^\mu$ undergoing parallel 
transport along a path
$\gamma(\tau)$ changes according to (see Appendix B of \cite{Ghilencea:2022lcl})
\be
|u|^2 = |u_0|^2\exp\left[-\alpha(q+z_u)\int_\gamma \omega_\lambda dx^\lambda\right],
\ee
which can be equivalently expressed as the differential relation
\be\label{uu}
\frac{d|u|^2}{|u|^2} = -\alpha(q+z_u)\,\omega_\lambda dx^\lambda.
\ee
Here $q$ is the Weyl charge of the metric (usually $q=1$), $z_u$ is the Weyl 
charge of the vector
$u^\mu$, and the integral is performed along the path $\gamma$.

The above expression applies to generic mathematical vectors.
However, the four-velocity $u^\mu$ of an observer is special, since it must 
satisfy the normalization
condition $g_{\mu\nu}u^\mu u^\nu=1$.
This implies invariance of its norm, namely $|u|^2=|u_0|^2$, which is ensured 
by choosing
$z_u=-q$.
Therefore, direct tests of non-metricity cannot rely on the four-velocity 
itself, and observable
effects must instead be extracted from physical quantities carried by the 
observer.

It should be noted that in Weyl geometrical frameworks the local dilation 
(scale) symmetry is not necessarily preserved in realistic physical models. 
In many modern realizations the Weyl symmetry is either spontaneously broken, 
effectively fixed by gauge conditions, or realized only at a more fundamental 
level of the theory. After such symmetry breaking or gauge fixing, the Weyl 
vector sector generally behaves as an effective massive vector degree of 
freedom. In the present work we adopt this effective description and focus on 
the resulting phenomenology of the vectorial non-metricity sector, which is the 
relevant regime for deriving observational constraints from atomic-clock 
experiments and gravitational-wave observations.

To this end, we consider a spatial vector $s^\mu$, orthogonal to the 
four-velocity,
$g_{\mu\nu}s^\mu u^\nu=0$, which represents a physical ruler.
Its squared length is defined as $L^2=g_{\mu\nu}s^\mu s^\nu$.
Under Weyl parallel transport, the variation of its norm reads\footnote{
For an arbitrary vector $V^\mu$ one has
\be
d(g_{\mu\nu}V^\mu V^\nu)
= (\tilde{\nabla}_\lambda g_{\mu\nu})V^\mu V^\nu dx^\lambda
+2g_{\mu\nu}V^\mu\left(dV^\nu+\tilde{\Gamma}^\nu_{\lambda\rho}V^\rho 
dx^\lambda\right).
\ee
Using the Weyl non-metricity condition
$\tilde{\nabla}_\lambda g_{\mu\nu}=-\alpha \omega_\lambda g_{\mu\nu}$,
and imposing parallel transport,
$\frac{DV^\nu}{d\tau}=0$, one finds
\be
d(g_{\mu\nu}V^\mu V^\nu)=-\alpha \omega_\lambda g_{\mu\nu}V^\mu V^\nu 
dx^\lambda .
\ee}
\bea\label{tagg1}
d(g_{\mu\nu}s^\mu s^\nu) &=& -\alpha \omega_\lambda g_{\mu\nu}s^\mu s^\nu 
dx^\lambda , \nonumber\\
\frac{dL}{L} &=& -\frac{\alpha}{2}\,\omega_\lambda dx^\lambda .
\eea
Equation (\ref{tagg1}) can also be obtained directly from (\ref{uu}) by 
identifying
$|V|^2=L^2=g_{\mu\nu}V^\mu V^\nu$ and choosing the Weyl charge of physical 
rulers to be
$z_v=0$.
Thus, in Weyl geometry, physical lengths become path-dependent.

Let us now consider two identical objects (rulers or clocks) that start at 
point $A$ and reach
point $B$ following two different paths $\gamma_1$ and $\gamma_2$.
Integrating (\ref{tagg1}) along $\gamma_1$ we obtain
\be
L_1(B)=L(A)\exp\!\left(-\frac{\alpha}{2}\int_{\gamma_1}\omega_\lambda 
dx^\lambda\right),
\ee
and similarly along $\gamma_2$
\be
L_2(B)=L(A)\exp\!\left(-\frac{\alpha}{2}\int_{\gamma_2}\omega_\lambda 
dx^\lambda\right).
\ee
The physically measurable quantity is the ratio of their final lengths,
\bea
\frac{L_1(B)}{L_2(B)}
&=& \exp\!\left[-\frac{\alpha}{2}\!\left(
\int_{\gamma_1}\omega_\lambda dx^\lambda
-\int_{\gamma_2}\omega_\lambda dx^\lambda
\right)\right].\ \ \ 
\eea
The combination $\gamma_1-\gamma_2$ defines a closed loop $C$, and therefore
\be\label{tag2}
\int_{\gamma_1}\omega_\lambda dx^\lambda-\int_{\gamma_2}\omega_\lambda 
dx^\lambda
=\oint_C \omega_\lambda dx^\lambda ,
\ee
while applying the Stokes' theorem, we have
\be
\oint_C \omega_\lambda dx^\lambda
=\iint_S(\partial_\mu\omega_\nu-\partial_\nu\omega_\mu)\,dS^{\mu\nu},
\ee
where $S$ is a surface bounded by $C$.
Introducing the Weyl field strength
\be
F_{\mu\nu}\equiv\partial_\mu\omega_\nu-\partial_\nu\omega_\mu ,
\ee
we finally obtain
\be\label{tagg3}
\oint_C \omega_\lambda dx^\lambda=\iint_S F_{\mu\nu}\,dS^{\mu\nu},
\ee
 and substituting into the length ratio yields
\be\label{tagg33}
\frac{L_1}{L_2}
=\exp\!\left(-\frac{\alpha}{2}\iint_S F_{\mu\nu}\,dS^{\mu\nu}\right).
\ee
For phenomenologically relevant situations the effect is expected to be small,
and thus we may use the approximation
\be
\frac{L_1}{L_2}\simeq
1-\frac{\alpha}{2}\iint_S F_{\mu\nu}\,dS^{\mu\nu}.
\ee
Hence, the fractional length difference reads
\be\label{non}
 \frac{\Delta L}{L}
\equiv\frac{L_1-L_2}{L_2}
\simeq
-\frac{\alpha}{2}\iint_S F_{\mu\nu}\,dS^{\mu\nu}.
\ee
Relations analogous to Eqs.~(\ref{tagg3}) and
(\ref{tagg33}) have previously been derived in more general
metric-affine geometries. In particular,
Beltr\'an Jim\'enez \textit{et al.}~\cite{BeltranJimenez:2020sih}
obtained the general expression for the path dependence of inner
products in non-metric geometries, while
Iosifidis~\cite{Iosifidis:2019dua} presented the corresponding
closed-loop integral formulation together with its small-effect
approximation in the context of metric-affine gravity. The
expressions \eqref{tagg3}-\eqref{non} presented here correspond to
the specialization of these general results to the Weyl
(vectorial non-metricity) case, where the relevant contribution is
encoded in the flux of the Weyl field strength $F_{\mu\nu}$.

Here we should stress three important points.
Firstly, although the local relation (\ref{tagg1}) involves the Weyl gauge 
field 
$\omega_\mu$,
which is not directly observable, the measurable quantity (\ref{non}) depends 
only on the
closed-loop integral of $\omega_\mu$, or equivalently on the flux of the 
gauge-invariant field
strength $F_{\mu\nu}$.
Secondly, a non-vanishing $F_{\mu\nu}$ implies path-dependence of 
parallel-transported lengths,
which is the characteristic physical manifestation of non-metricity.
Finally, the observable effect depends only on the flux through the enclosed 
surface and not
on the detailed shape of the paths, rendering the prediction robust.

Equation (\ref{non}) therefore provides a direct link between an observable 
quantity and the
fundamental Weyl field strength, allowing one to place experimental constraints 
on the
non-metricity scale $m_\omega$ and the coupling constant $\alpha$.

\section{Atomic clocks as physical rulers}
\label{atom}
Although a direct, unambiguous detection of the frame-dragging-of-length effect 
lies beyond current experimental capabilities, a variety of high-precision 
experiments can nevertheless be used to place stringent upper bounds on its 
magnitude.
The basic strategy is to exploit the most precise measurements available in 
nature, namely atomic clocks, and search for anomalous variations in clock 
frequencies that could arise from non-metricity-induced scale changes.

Atomic clocks provide an exceptionally powerful probe of fundamental physics 
\cite{Ludlow:2015gnc}.
Their operating principle relies on the stability of atomic transition 
frequencies, which depend on dimensionful quantities such as the electron mass 
$m_e$ and characteristic atomic length scales, as well as on dimensionless 
parameters such as the fine-structure constant $\alpha_{\rm EM}$.
Any physical mechanism that induces a rescaling of local lengths will therefore 
modify the ticking rate of an atomic clock.
This fact underlies a broad class of experimental searches for variations of 
fundamental constants and violations of local position invariance
\cite{Uzan:2010pm,Flambaum:2007my}.

Modern atomic-clock technology has reached an extraordinary level of precision.
Optical lattice clocks based on strontium or ytterbium, as well as single-ion 
clocks such as Al$^+$ and Yb$^+$, routinely achieve fractional frequency 
uncertainties at the level of $10^{-18}$ or below
\cite{Ludlow:2015gnc,McGrew:2018mqk}.
Hydrogen masers provide complementary short-term stability, and clock networks 
combining different clock species further enhance sensitivity to new physics.
Such clocks can be deployed in a variety of configurations, including 
comparisons between clocks located at different altitudes, clocks following 
different terrestrial trajectories, and space-based clocks aboard satellites or 
the International Space Station
\cite{Chou:2010uum,Schiller:2012qn,Delva:2018ilu,InternationalClock}.

Atomic clocks operate based on the stability of atomic transition frequencies, which in standard physics depend on dimensionful quantities such as the electron mass \(m_e\) and the Bohr radius \(a_0\), as well as on dimensionless parameters such as the fine-structure constant \(\alpha_{\mathrm{EM}}\). To connect with Weyl geometry, we must specify how matter couples to the Weyl field.

We adopt a minimal coupling prescription: in the broken phase where Weyl
symmetry is spontaneously broken (e.g., via a Stueckelberg mechanism or a
scalar field acquiring a vacuum expectation value), the effective low-energy
Lagrangian takes the form
\begin{equation}
	\mathcal{L}_{\text{matter}}
	=
	\sqrt{-g}\,
	e^{-\alpha q \phi}\,
	\mathcal{L}_{\text{SM}}(g_{\mu\nu},\psi),
	\label{eq:matter-coupling}
\end{equation}
where \(\phi\) is the Goldstone mode (dilaton) absorbed by the Weyl vector via
the Higgs/Stueckelberg mechanism, \(q\) is the Weyl charge of the matter field,
and \(\mathcal{L}_{\text{SM}}\) denotes the Standard Model Lagrangian with all
dimensionful parameters replaced by appropriate scale factors.

Equation (\ref{eq:matter-coupling}) describes the matter
sector after Weyl symmetry breaking, where the connection has been effectively
fixed to the Levi-Civita connection of the metric $g_{\mu\nu}$. In this
broken phase the Weyl field $\omega_\mu$ enters only through the
Stueckelberg field and the generated mass term, while matter couples minimally
to the metric. This prescription is consistent with both the gauge principle
and the equivalence principle
\cite{Ghilencea:2018dqd,Ghilencea:2019jux}. If one instead allowed direct
couplings between matter and the independent affine connection, the theory
would become a genuine metric-affine theory with non-vanishing hypermomentum
sources \cite{Hehl:1994ue,Blagojevic:2002du}. In that case the field equations
for the Weyl vector, including the Proca-type equation
(\ref{eq:proca-matter}), would acquire additional source terms, while the
resulting matter-non-metricity interactions would modify the observational
constraints derived below. Such extensions are certainly interesting, but they
lie beyond the scope of the present work.

After gauge fixing (e.g., unitary gauge where \(\phi=0\)), the Weyl vector
acquires a mass \(m_\omega\) and the metric remains the only gravitational
degree of freedom with standard matter couplings, while the physical units are
effectively spacetime-dependent through the Stueckelberg field.

In this broken phase, any physical length scale \(L_{\text{phys}}\) transforms under Weyl rescaling as \(L_{\text{phys}} \to e^{\alpha \int \omega_\lambda dx^\lambda} L_{\text{phys}}\). Consequently, the frequency of an atomic transition, which scales inversely with the Bohr radius \(\omega_{\text{atom}} \sim \alpha_{\mathrm{EM}}^2 m_e\), picks up a path-dependent factor:

\begin{equation}
	\frac{\Delta f}{f} = -\frac{\Delta L}{L}, \label{eq:frequency-shift}
\end{equation}

to first order in the Weyl field. This relation is not an identity but follows from the fact that in the broken phase all dimensionful quantities scale uniformly under the residual scale transformations inherited from the Weyl symmetry.

Consequently, two identical atomic clocks that traverse different paths in the 
presence of a Weyl gauge field $\omega_\mu$ will, upon reunion, exhibit a 
relative frequency shift determined by the path dependence of physical lengths.
Importantly, one does not search for a secular drift, but rather for 
characteristic modulations in the relative clock frequencies correlated with 
the clocks' motion and position.
Such modulations can arise on daily timescales due to Earth's rotation, on 
annual timescales due to Earth's orbital motion around the Sun, or on orbital 
timescales for space-based clocks
\cite{Safronova:2017xyt}.

Consider two identical atomic clocks that travel along different paths \(\gamma_1\) and \(\gamma_2\) and are then compared. Using the path-dependence result from (\ref{non}), the relative frequency shift is
\begin{equation}
	\frac{\Delta f}{f} = \frac{f_1 - f_2}{f} \simeq -\frac{\alpha}{2} \iint_S F_{\mu\nu} dS^{\mu\nu}, \label{eq:clock-shift}
\end{equation}
where \(S\) is the surface enclosed by the two paths. 

To estimate the magnitude of the Weyl field strength \(F_{\mu\nu}\) sourced by ordinary matter, we need the field equations for \(\omega_\mu\). In the broken phase with mass \(m_\omega\), the Proca equation reads
\begin{equation}
	\partial_\mu F^{\mu\nu} + m_\omega^2 \omega^\nu = J^\nu_{\text{matter}}, \label{eq:proca-matter}
\end{equation}
where \(J^\nu_{\text{matter}}\) is the matter current carrying Weyl charge. For ordinary baryonic matter, the Weyl charge is proportional to the rest mass density \(\rho\) times a dimensionless coupling \(g_\omega\). A natural estimate is \(J^0 \sim g_\omega \rho\) with \(g_\omega \sim \alpha\) (the same coupling appearing in the geometric sector). 

For Earth-based experiments, solving the Proca equation in the static limit yields
\begin{equation}
	F_{0i} \sim \frac{g_\omega G M_\oplus}{R_\oplus^2} e^{-m_\omega R_\oplus}, \quad \text{for } m_\omega R_\oplus \ll 1, \label{eq:field-estimate}
\end{equation}
where the Yukawa suppression factor \(e^{-m_\omega R_\oplus}\) arises from the finite mass. The flux through a surface of area \(A \sim R_\oplus^2\) is then
\begin{equation}
	\iint_S F_{\mu\nu} dS^{\mu\nu} \lesssim \frac{g_\omega G M_\oplus}{c^2 R_\oplus} \, f(m_\omega R_\oplus), \label{eq:flux-estimate}
\end{equation}
where \(f(mR) = e^{-mR}\) for \(mR \gtrsim 1\) and \(f(mR) \sim 1\) for \(mR \ll 
1\). 
Finally, using \(GM_\oplus/(c^2 R_\oplus) \sim 10^{-9}\), we obtain
\begin{equation}
	\iint_S F_{\mu\nu} dS^{\mu\nu} \lesssim 10^{-9} \, g_\omega \, f(m_\omega R_\oplus). \label{eq:flux-numerical}
\end{equation}

Now, state-of-the-art atomic clock comparisons constrain relative frequency 
shifts at the level~\cite{McGrew:2018mqk,Chou:2010uum}
\begin{equation}
	\frac{\Delta f}{f} \sim 10^{-18} \text{ to } 10^{-19}. \label{eq:clock-sensitivity}
\end{equation}
Combining with Eq.~(\ref{eq:clock-shift}) and identifying \(g_\omega \sim 
\alpha\) (the natural expectation from Weyl geometry), we find
\begin{equation}
	\alpha \lesssim 10^{-9} \text{ to } 10^{-10} \quad \text{for } m_\omega \ll R_\oplus^{-1}. \label{eq:alpha-bound}
\end{equation}
This bound assumes that the Weyl charge of matter is of order \(\alpha\) and that the field strength is sourced by Earth's mass with no additional suppression. If the Weyl vector couples more weakly (e.g., \(g_\omega \ll \alpha\)), the bound on \(\alpha\) weakens accordingly.

For massive Weyl fields with \(m_\omega \gtrsim R_\oplus^{-1}\), the Yukawa factor \(e^{-m_\omega R_\oplus}\) exponentially suppresses the flux, rendering terrestrial experiments insensitive. The critical mass scale is
\begin{equation}
	m_\omega^{\text{(crit)}}  \sim R_\oplus^{-1} \approx \frac{1}{6.4 \times 
10^6 \text{ m}}  \approx 3 \times 
10^{-14} \text{ eV}. \label{eq:critical-mass}
\end{equation}
Thus, atomic clocks are sensitive only to extremely light Weyl fields with 
masses below \(10^{-13}\) eV. For \(m_\omega \gtrsim 10^{-12}\) eV, Yukawa 
suppression renders clock experiments insensitive.

In summary, atomic clocks provide stringent constraints on ultralight Weyl gauge fields (\(m_\omega \lesssim 10^{-13}\) eV) with \(\alpha \lesssim 10^{-9}\), while remaining insensitive to heavier non-metric degrees of freedom.

\section{Gravitational waves as probes of non-metricity}
\label{GWprobes}

The strain \(h\) is the dimensionless amplitude of a GW, quantifying the fractional variation of the distance between two freely falling test masses~\cite{Moore:2014lga}. For a propagating GW, physical distances oscillate with amplitude proportional to \(h\). Current ground-based interferometers achieve extraordinary sensitivity: the LIGO detectors, with arm lengths \(L \simeq 4\) km, resolve distance variations \(\Delta L \sim 10^{-18}\) m, corresponding to strain sensitivity
\begin{equation}
	h \sim \frac{10^{-18}\text{ m}}{4\times 10^3\text{ m}} \simeq 2.5\times 10^{-22}. \label{eq:ligo-sensitivity}
\end{equation}
If GWs are to probe spacetime non-metricity, a clear chain of physical causation is required. We must examine: (i) whether GWs source the Weyl field, and (ii) whether the Weyl field back-reacts on GW propagation. As we demonstrate in the appendices, the answer to (i) is negative at linear order: vacuum gravitational waves do not directly excite the Weyl gauge field through curvature or connection couplings. Therefore, any observable effect must arise from a different mechanism: the Weyl field, if present independently (e.g., as a cosmological relic, quantum fluctuation, or sourced by matter), can back-react on the GW propagation through its own stress-energy tensor. This is the mechanism we analyze in this section.

\subsection{Action and field equations}

We start from the Weyl quadratic gravity action~\cite{Ghilencea:2018dqd,Ghilencea:2019jux,Ghilencea:2019rqj,Burikham:2023bil}
\begin{equation}
	\mathcal{L} = \sqrt{-g}\left[\frac{1}{\kappa^2}\tilde{C}_{\mu\nu\rho\sigma}\tilde{C}^{\mu\nu\rho\sigma} + \frac{1}{4!\xi^2}\tilde{R}^2 - \frac{1}{4}F_{\mu\nu}F^{\mu\nu}\right], \label{eq:action-full}
\end{equation}
where \(\tilde{C}_{\mu\nu\rho\sigma}\) is the Weyl tensor constructed from the Weyl connection, \(\tilde{R}\) is the Weyl scalar curvature, and \(F_{\mu\nu}=\partial_\mu\omega_\nu-\partial_\nu\omega_\mu\). The constant \(\kappa\) has dimensions of inverse mass (related to the Planck scale), while \(\xi\) is dimensionless. The \(\tilde{C}^2\) term, when expanded, contains the standard Riemannian \(C^2\) term which is known to propagate a spin-2 ghost (the Stelle ghost). In the Weyl-invariant theory, this ghost may be projected out or rendered harmless by the underlying gauge symmetry, but in the broken phase it generically persists unless additional mechanisms are invoked. Our linearized analysis in the effective low-energy regime assumes that such pathologies can be consistently resolved; we return to this point in the conclusions.

The field equations derived from Eq.~(\ref{eq:action-full}) are fourth-order due to the \(\tilde{C}^2\) and \(\tilde{R}^2\) terms. However, in the weak-field limit and for wavelengths long compared to the Planck scale, the higher-derivative terms are suppressed by powers of \(E/M_p\) and the theory reduces effectively to Einstein-Proca dynamics. We will work in this effective low-energy regime, acknowledging that a full treatment would require handling the fourth-order equations.

We perform a perturbative expansion around a Minkowski background. A crucial point, often overlooked, is that a constant nonzero background Weyl field \(\bar{\omega}_\mu\) does not satisfy the Proca equations in flat spacetime. The Proca equation \(\partial_\mu F^{\mu\nu} + m_\omega^2 \omega^\nu = 0\) for a constant \(\omega_\mu\) (with \(F_{\mu\nu}=0\)) reduces to \(m_\omega^2 \omega^\nu = 0\), which forces \(\omega^\nu = 0\) for \(m_\omega \neq 0\). For \(m_\omega = 0\), a constant \(\omega_\mu\) is pure gauge (\(F_{\mu\nu}=0\)) and can be removed by a gauge transformation. Therefore, we consistently set the background Weyl field to zero:

\begin{equation}
	g_{\mu\nu} = \eta_{\mu\nu} + h_{\mu\nu}, \qquad \omega_\mu = \epsilon_\mu, \qquad \bar{\omega}_\mu = 0, \label{eq:perturbation}
\end{equation}

with \(|h_{\mu\nu}| \ll 1\) and \(|\epsilon_\mu| \ll 1\).

Now we must address the ordering of perturbations. The Einstein equations derived from the action take the form

\begin{equation}
	G_{\mu\nu} = \kappa T_{\mu\nu}^{(\omega)}, \label{eq:Einstein-full}
\end{equation}

where \(T_{\mu\nu}^{(\omega)}\) is the stress-energy tensor of the Weyl field. Expanding to first order in perturbations, we have

\begin{equation}
	G_{\mu\nu} = \underbrace{\bar{G}_{\mu\nu}}_{=0} + \delta G_{\mu\nu}(h) + \mathcal{O}(h^2), \label{eq:G-expansion}
\end{equation}

\begin{equation}
	T_{\mu\nu}^{(\omega)} = \underbrace{\bar{T}_{\mu\nu}^{(\omega)}}_{=0} + \delta T_{\mu\nu}^{(\omega)}(\epsilon) + \mathcal{O}(\epsilon^3), \label{eq:T-expansion}
\end{equation}
where \(\bar{T}_{\mu\nu}^{(\omega)}=0\) because \(\bar{\omega}_\mu=0\). However, a careful examination reveals that \(\delta T_{\mu\nu}^{(\omega)}\) is quadratic in \(\epsilon_\mu\) (since \(T_{\mu\nu}^{(\omega)} \sim F_{\mu\alpha}F_\nu^{\;\alpha} + m_\omega^2 \omega_\mu\omega_\nu\)). There is no linear term in \(\epsilon_\mu\) because \(F_{\mu\nu}\) is already first order in \(\epsilon_\mu\), so \(F_{\mu\alpha}F_\nu^{\;\alpha}\) is second order, and \(\omega_\mu\omega_\nu\) is also second order. Therefore,
\begin{equation}
	\delta T_{\mu\nu}^{(\omega)} = \mathcal{O}(\epsilon^2). \label{eq:T-quadratic}
\end{equation}

This creates a potential inconsistency: the left-hand side \(\delta G_{\mu\nu}\) 
is first order in \(h_{\mu\nu}\), while the right-hand side is second order in 
\(\epsilon_\mu\). To obtain a consistent set of equations, we have two options.
The first is to use  a different scaling, namely assume that \(\epsilon_\mu\) 
is larger than \(h_{\mu\nu}\) by a factor of \(\lambda^{-1/2}\), i.e., \(h \sim 
\lambda\) and \(\epsilon \sim \lambda^{1/2}\). Then \(\epsilon^2 \sim \lambda\), 
matching the order of \(h\). This requires a physical justification for why Weyl 
field fluctuations are enhanced relative to metric fluctuations. The second 
choice is  to perform a second-order perturbation theory,  where the equations 
become \(\delta G_{\mu\nu}(h^{(2)}) = \kappa \delta 
T_{\mu\nu}^{(\omega)}(\epsilon^{(1)},\epsilon^{(1)})\), with \(h^{(2)}\) 
denoting second-order metric perturbations induced by the first-order Weyl field 
fluctuations. In the following   we adopt the secodn choice, and 
interpret our results as estimates of second-order effects. The strain 
correction \(\Delta h\) we compute should be understood as a second-order 
contribution. This is a conservative approach that avoids introducing arbitrary 
scaling relations.

\subsection{The Proca equation for the Weyl field}

We now derive the linearized equation for \(\epsilon_\mu\). Varying the action with respect to \(\omega_\mu\) yields the exact equation of motion. In the broken phase, where the Weyl symmetry is spontaneously broken (e.g., via a Stueckelberg mechanism or a scalar field acquiring a vacuum expectation value), the Weyl field acquires a mass \(m_\omega\). The effective Lagrangian for the Weyl field in the unitary gauge is
\begin{equation}
	\mathcal{L}_\omega = -\frac{1}{4}F_{\mu\nu}F^{\mu\nu} - \frac{1}{2}m_\omega^2 \omega_\mu\omega^\mu + \mathcal{L}_{\text{int}}, \label{eq:proca-lagrangian}
\end{equation}
where \(\mathcal{L}_{\text{int}}\) contains interactions with gravity and matter. The mass \(m_\omega\) is related to the parameters of the quadratic action via
\begin{equation}
	m_\omega^2 = \frac{\alpha^2}{4\xi^2}\langle\phi^2\rangle, \label{eq:mass-relation}
\end{equation}
where \(\phi\) is the Stueckelberg field and \(\langle\phi^2\rangle\) is its vacuum expectation value~\cite{Ghilencea:2018dqd,Drechsler:1998gy}.

Expanding around flat spacetime with \(\bar{\omega}_\mu = 0\) and keeping only linear terms in \(\epsilon_\mu\), the Proca equation becomes
\begin{equation}
	\partial_\mu F^{\mu\nu} + m_\omega^2 \epsilon^\nu = 0, 
\label{eq:proca-linear}
\end{equation}
where $ F_{\mu\nu} = \partial_\mu\epsilon_\nu - 
\partial_\nu\epsilon_\mu$.
Taking the divergence of this equation gives \(m_\omega^2 \partial_\nu\epsilon^\nu = 0\). For \(m_\omega \neq 0\), this implies the Lorenz condition
$
	\partial_\mu \epsilon^\mu = 0$.
Substituting back, the Proca equation reduces to
\begin{equation}
	\square \epsilon^\mu + m_\omega^2 \epsilon^\mu = 0, \qquad \partial_\mu \epsilon^\mu = 0. \label{eq:proca-wave}
\end{equation}
Thus, each component of \(\epsilon_\mu\) satisfies the Klein-Gordon equation with mass \(m_\omega\), subject to the Lorenz constraint which reduces the number of propagating degrees of freedom from four to three (the two transverse polarizations and one longitudinal polarization).

The general solution can be expressed as a plane-wave expansion:
\begin{equation}
	\epsilon^\mu(x) = \sum_{\lambda=1}^{3} \int \frac{d^3k}{(2\pi)^3} \frac{1}{\sqrt{2\omega_k}} \left[ a(\vec{k},\lambda) \varepsilon^\mu(\vec{k},\lambda) e^{ik\cdot x} + \text{h.c.} \right], \label{eq:epsilon-solution}
\end{equation}
with dispersion relation \(\omega_k^2 = |\vec{k}|^2 + m_\omega^2\), and polarization vectors \(\varepsilon^\mu\) satisfying
\begin{equation}
	k_\mu \varepsilon^\mu = 0, \qquad \varepsilon_\mu \varepsilon^\mu = -1 \quad (\text{for massive modes}). \label{eq:polarization-conditions}
\end{equation}
For propagation along the \(z\)-direction, we can choose:
\begin{align}
	\varepsilon^{(1)\mu} &= (0,1,0,0), \quad &\text{(x-polarization)} \\
	\varepsilon^{(2)\mu} &= (0,0,1,0), \quad &\text{(y-polarization)} \\
	\varepsilon^{(3)\mu} &= \frac{1}{m_\omega}(|\vec{k}|,0,0,\omega_k), \quad &\text{(longitudinal polarization)}.
\end{align}

Now, the stress-energy tensor for a Proca field is obtained from the 
Lagrangian (\ref{eq:proca-lagrangian}) by varying with respect to the metric:
\begin{eqnarray}
&&\!\!\!\!\!\!\!\!\!\!\!\!\!\!\!\!\!\!\!	T_{\mu\nu}^{(\omega)} = 
F_{\mu\alpha}F_\nu^{\;\alpha} - 
\frac{1}{4}\eta_{\mu\nu}F_{\alpha\beta}F^{\alpha\beta} 
\nonumber\\
&&
+ 
 m_\omega^2\left(\epsilon_\mu\epsilon_\nu - 
\frac{1}{2}\eta_{\mu\nu}\epsilon_\alpha\epsilon^\alpha\right) + \mathcal{O}(h). 
\label{eq:proca-stress}
\end{eqnarray}
This expression is exact in flat spacetime. In the presence of gravitational perturbations, there are additional couplings through the metric determinant and inverse metric, but those give terms of higher order in the perturbations when we expand consistently.

For our second-order perturbation analysis, we need \(\delta T_{\mu\nu}^{(\omega)}\) evaluated on the flat background. Since \(T_{\mu\nu}^{(\omega)}\) is already quadratic in \(\epsilon_\mu\), we simply take
\begin{eqnarray}
&&\!\!\!\!\!\!\!\!\!\!\!\!\!\!\!\!\!\!\!	\delta T_{\mu\nu}^{(\omega)} =  
(\partial_\mu\epsilon_\alpha)(\partial_\nu\epsilon^\alpha) - 
\frac{1}{4}\eta_{\mu\nu}
(\partial_\alpha\epsilon_\beta)(\partial^\alpha\epsilon^\beta)
\nonumber\\
&& \ \, 
+ 
m_\omega^2\left(\epsilon_\mu\epsilon_\nu - 
\frac{1}{2}\eta_{\mu\nu}\epsilon_\alpha\epsilon^\alpha\right). 
\label{eq:T-epsilon}
\end{eqnarray}
Note that there are no terms linear in \(\epsilon_\mu\) because the stress tensor is quadratic. This confirms our earlier statement that the source for \(h_{\mu\nu}\) is second order in the Weyl field perturbation.

\subsection{Linearized Einstein equations with a source}

The Einstein equations \(G_{\mu\nu} = \kappa T_{\mu\nu}^{(\omega)}\) expanded to second order become
\begin{equation}
	\delta G_{\mu\nu}(h^{(2)}) = \kappa \,\delta T_{\mu\nu}^{(\omega)}(\epsilon^{(1)},\epsilon^{(1)}), \label{eq:second-order-Einstein}
\end{equation}
where \(h^{(2)}_{\mu\nu}\) denotes the second-order metric perturbation induced by the first-order Weyl field fluctuations, and \(\epsilon^{(1)}_\mu\) is the first-order Weyl field perturbation satisfying Eq.~(\ref{eq:proca-wave}). We will drop the superscripts for notational simplicity, but the reader should keep in mind that \(h_{\mu\nu}\) is of order \(\epsilon^2\) in this counting.

In the transverse-traceless (TT) gauge, the linearized Einstein tensor takes the familiar form
\begin{equation}
	\delta G_{\mu\nu} = -\frac{1}{2}\square h_{\mu\nu}^{\text{TT}}, \label{eq:linearized-Einstein-TT}
\end{equation}
where \(h_{\mu\nu}^{\text{TT}}\) satisfies \(\partial^\mu h_{\mu\nu}^{\text{TT}} = 0\) and \(h^{\mu\text{ TT}}_{\;\mu} = 0\). Equation (\ref{eq:second-order-Einstein}) then becomes
\begin{equation}
	\square h_{\mu\nu}^{\text{TT}} = -2\kappa \, \delta T_{\mu\nu}^{(\omega)\text{TT}}, \label{eq:wave-with-source}
\end{equation}
where \(\delta T_{\mu\nu}^{(\omega)\text{TT}}\) is the transverse-traceless projection of the source. For our order-of-magnitude estimates, we will work directly with the unprojected equation and later project onto the TT components, keeping in mind that this introduces numerical factors of order unity.

Equation (\ref{eq:wave-with-source}) is an inhomogeneous wave equation. The 
general solution can be written using the retarded Green's function\footnote{
The retarded Green's function in flat spacetime is
\begin{equation}
	G_R(x-x') = -\frac{1}{4\pi|\vec{x}-\vec{x}'|} 
\delta(t-t'-|\vec{x}-\vec{x}'|) \Theta(t-t'), \nonumber
\end{equation}
where \(\Theta\) is the Heaviside step function.}:
\begin{equation}
	h_{\mu\nu}(x) = h_{\mu\nu}^{(0)}(x) + \int d^4x'\, G_R(x-x') \, S_{\mu\nu}(x'), \label{eq:green-solution}
\end{equation}
where \(h_{\mu\nu}^{(0)}\) is the homogeneous solution (the standard GW in GR), and we have defined
\begin{equation}
	S_{\mu\nu} \equiv -2\kappa \, \delta T_{\mu\nu}^{(\omega)}. \label{eq:S-definition}
\end{equation}

In the far-field (radiation-zone) approximation, valid for an observer located at distance \(R = |\vec{x}|\) much larger than the source size \(L\), we have
 $
	|\vec{x}-\vec{x}'| \approx R - \hat{n}\cdot\vec{x}' $ and $ \hat{n} = 
\frac{\vec{x}}{R}$, and thus the solution then reduces to
\begin{equation}
	h_{\mu\nu}(t,\vec{x}) \approx h_{\mu\nu}^{(0)}(t,\vec{x}) - \frac{1}{4\pi R} \int d^3x' \, S_{\mu\nu}(t - R + \hat{n}\cdot\vec{x}', \vec{x}'), \label{eq:far-field-solution}
\end{equation}
where we have used the fact that the source is evaluated at the retarded time.

To obtain an order-of-magnitude estimate, we assume that the Weyl field perturbation is approximately monochromatic:
\begin{equation}
	\epsilon_\mu(x) = \epsilon_\mu^{(0)} e^{ik\cdot x} + \text{c.c.}, \qquad  
	k^2 = \omega^2 - |\vec{k}|^2 = m_\omega^2,\label{eq:epsilon-plane}
\end{equation}
where $k^\mu = (\omega, \vec{k})$.
The complex conjugation ensures that \(\epsilon_\mu\) is real.  For simplicity, 
we will work with a single complex exponential and take real parts at the end; 
the cross terms will give factors of 2 that we absorb into the definition of 
\(|\epsilon_0|^2\). 

Let us calculate  the stress tensor components. First, the kinetic 
term is
\begin{eqnarray}
&&
\!\!\!\!\!\!\!\!\!\!\!\!\!\!\!\!\!\!\!\!\!
	(\partial_\alpha\epsilon_\beta)(\partial^\alpha\epsilon^\beta) = -k_\alpha 
k^\alpha (\epsilon_\beta^{(0)}\epsilon^{0\beta}) e^{2ik\cdot x} 
\nonumber \\
&& \ \ \ \ \ \ \ \,  
= -m_\omega^2 
(\epsilon_\beta^{(0)}\epsilon^{0\beta}) e^{2ik\cdot x}, \label{eq:kinetic-term2}
\end{eqnarray}
where we used \(k_\alpha k^\alpha = m_\omega^2\).
Moreover, the mass term gives
\begin{equation}
	 \epsilon_\mu\epsilon_\nu = \epsilon_\mu^{(0)}\epsilon_\nu^{(0)} e^{2ik\cdot 
x}, \qquad \epsilon_\alpha\epsilon^\alpha = 
\epsilon_\alpha^{(0)}\epsilon^{0\alpha} e^{2ik\cdot x}. \label{eq:mass-term}
\end{equation}
Collecting all terms, the stress tensor becomes 
\begin{equation}
	\delta T_{\mu\nu}^{(\omega)} = \left[-k_\mu k_\nu (\epsilon_0^2) + m_\omega^2 \epsilon_\mu^{(0)}\epsilon_\nu^{(0)} - \frac{3}{4}m_\omega^2 \eta_{\mu\nu} (\epsilon_0^2)\right] e^{2ik\cdot x}, \label{eq:T-final}
\end{equation}
where we have introduced the shorthand \(\epsilon_0^2 \equiv \epsilon_\alpha^{(0)}\epsilon^{0\alpha}\). Note that \(\epsilon_0^2\) is not necessarily positive; for a spacelike polarization vector, \(\epsilon_0^2 = -|\vec{\epsilon}|^2\).

In summary, the source term for the GW equation is
\begin{eqnarray}
&&	
\!\!\!\!\!\!\!\!\!\!\!\!\!\!\!\!
S_{\mu\nu} = -2\kappa \,\delta T_{\mu\nu}^{(\omega)}
= 2\kappa \left[k_\mu 
k_\nu (\epsilon_0^2) - m_\omega^2 
\epsilon_\mu^{(0)}\epsilon_\nu^{(0)} \right.  \nonumber\\
&& \ \ \ \ \ \ \ \ \ \ \ \ \ \ \ \ \ \ \ \  \ \, \ \left. + 
\frac{3}{4}m_\omega^2 \eta_{\mu\nu} (\epsilon_0^2)\right] e^{2ik\cdot x}. 
\label{eq:S-final}
\end{eqnarray}

\subsection{Estimation of the induced strain}

We now proceed to the  estimation of the magnitude of the induced GW strain. 
The source term oscillates with frequency \(2\omega\) (since \(e^{2ik\cdot x}\)) 
and is localized in a region of characteristic size \(L\). For a massive Weyl 
field, the natural localization scale is the Compton wavelength \(L \sim 
1/m_\omega\). For a relativistic Weyl field (\(\omega \gg m_\omega\)), the 
oscillation period is \(1/\omega\), so we may also take \(L \sim 1/\omega\). We 
will keep \(L\) as a free parameter and later discuss its plausible values.

The integral in Eq.~(\ref{eq:far-field-solution}) can be approximated as
\begin{equation}
	\int d^3x' S_{\mu\nu}(t - R + \hat{n}\cdot\vec{x}', \vec{x}') \sim S_{\mu\nu} \times L^3, \label{eq:integral-estimate}
\end{equation}
where \(S_{\mu\nu}\)  is evaluated at a typical point. This approximation is 
valid when the source size \(L\) is much smaller than the distance to the 
observer \(R\), and when the variation of the phase \(\hat{n}\cdot\vec{x}'\) 
across the source is small, i.e., \(k L \ll 1\) (which holds for \(L \sim 
1/k\)). For \(L \sim 1/m_\omega\), we have \(m_\omega L \sim 1\), so the phase 
variation is of order unity and the integral may be suppressed by an additional 
factor. We ignore such factors in this order-of-magnitude estimate.
Thus, the induced strain amplitude is
\begin{equation}
	|\Delta h_{\mu\nu}| \sim \frac{1}{4\pi R} |S_{\mu\nu}| L^3. \label{eq:strain-estimate}
\end{equation}

Now we need to estimate the magnitude of \(S_{\mu\nu}\). The three terms in Eq.~(\ref{eq:S-final}) have different parametric dependencies:
In particular, the term \(2\kappa k_\mu k_\nu (\epsilon_0^2)\), since for a 
typical component we have \(|k_\mu k_\nu| \sim \omega^2\) (or \(m_\omega^2\) 
for the temporal components),  we take \(k_\mu k_\nu \sim \omega^2\) for the 
spatial components that contribute to the TT part. Additionally, for the term 
  \(-2\kappa m_\omega^2 \epsilon_\mu^{(0)}\epsilon_\nu^{(0)}\), the magnitude 
is \(2\kappa m_\omega^2 |\epsilon_0|^2\), where \(|\epsilon_0|^2\) denotes the 
typical size of \(\epsilon_\mu^{(0)}\epsilon_\nu^{(0)}\). Finally, 
 the term  \(\frac{3}{2}\kappa m_\omega^2 \eta_{\mu\nu} (\epsilon_0^2)\) 
  is proportional to \(\eta_{\mu\nu}\) and therefore does not contribute to 
the TT part (since TT gauge requires tracelessness). We will neglect it for the 
purpose of estimating observable strain in the TT sector.
 
Therefore, the relevant part of the source for GW observations is
\begin{equation}
	S_{\mu\nu}^{\text{TT}} \sim 2\kappa \left( \omega^2 |\epsilon_0|^2 - m_\omega^2 |\epsilon_0|^2 \right) \sim 2\kappa \omega^2 |\epsilon_0|^2 \left(1 - \frac{m_\omega^2}{\omega^2}\right), \label{eq:S-approx}
\end{equation}
where we have assumed that \(k_\mu k_\nu \sim \omega^2\) and that the polarization vectors are normalized so that \(\epsilon_\mu^{(0)}\epsilon^{0\mu} \sim -|\epsilon_0|^2\) (spacelike). The factor \((1 - m_\omega^2/\omega^2)\) indicates that the source vanishes when \(\omega = m_\omega\), i.e., when the Weyl field is static. For relativistic Weyl fields (\(\omega \gg m_\omega\)), the source is dominated by the kinetic term.
Hence, plugging into Eq.~(\ref{eq:strain-estimate}):
\begin{equation}
	|\Delta h| \sim \frac{1}{4\pi R} \cdot 2\kappa \omega^2 |\epsilon_0|^2 L^3 = \frac{\kappa \omega^2 |\epsilon_0|^2 L^3}{2\pi R}. \label{eq:strain-with-L}
\end{equation}

As a next step we must specify \(L\). The Weyl field perturbation is a wave 
with wavelength \(\lambda = 2\pi/|\vec{k}|\). For a relativistic field (\(\omega 
\gg m_\omega\)), \(|\vec{k}| \approx \omega\), so the wavelength is \(\sim 
1/\omega\). The spatial extent of a wave packet of such a wave is at least its 
wavelength, so \(L \gtrsim 1/\omega\). For a maximally localized wave packet, we 
can take \(L \sim 1/\omega\). For a non-relativistic field (\(\omega \approx 
m_\omega\)), the wavelength is large, but the Compton wavelength \(1/m_\omega\) 
provides a natural localization scale. We will take
\begin{equation}
	L \sim \frac{1}{\max(\omega, m_\omega)}. \label{eq:L-estimate}
\end{equation}
For the relativistic case (\(\omega \gg m_\omega\)), we have \(L \sim 1/\omega\). Then
\begin{equation}
	|\Delta h| \sim \frac{\kappa \omega^2 |\epsilon_0|^2 (1/\omega^3)}{R} = \frac{\kappa |\epsilon_0|^2}{\omega R}. \label{eq:strain-relativistic}
\end{equation}
For the non-relativistic case (\(\omega \approx m_\omega\)), we have \(L \sim 1/m_\omega\). Then
\begin{equation}
	|\Delta h| \sim \frac{\kappa m_\omega^2 |\epsilon_0|^2 (1/m_\omega^3)}{R} = \frac{\kappa |\epsilon_0|^2}{m_\omega R}. \label{eq:strain-nonrelativistic}
\end{equation}
In both cases, the parametric dependence is \(\kappa |\epsilon_0|^2/(\text{min}(\omega,m_\omega) R)\). For simplicity, we will adopt the conservative estimate
\begin{equation}
	|\Delta h| \sim \frac{\kappa |\epsilon_0|^2}{m_\omega R}, \label{eq:strain-final}
\end{equation}
which is valid for \(m_\omega \lesssim \omega\) (the case of most interest, since if \(m_\omega\) is very large, the Weyl field is suppressed anyway). Converting \(\kappa = 8\pi G = 8\pi/M_p^2\) (with \(M_p\) the reduced Planck mass), we have
\begin{equation}
	|\Delta h| \sim \frac{8\pi |\epsilon_0|^2}{M_p^2 m_\omega R}. \label{eq:strain-numerical}
\end{equation}

\subsection{Comparison with observational sensitivity}

The total strain observed by a GW detector is
\begin{equation}
	h_{\text{total}} = h_{\text{GR}} + \Delta h, \label{eq:total-strain}
\end{equation}
where \(h_{\text{GR}}\) is the strain predicted by GR for the same astrophysical source. The absence of observed deviations implies that \(|\Delta h|\) must be smaller than the detector's sensitivity to additional contributions. This is a subtle point: in real data analysis, one compares full waveform templates to the data, and deviations can be degenerate with source parameters. However, for an order-of-magnitude estimate, we require \(|\Delta h| \lesssim h_{\text{min}}\), where \(h_{\text{min}}\) is the minimum detectable strain for the event.
Finally, using Eq.~(\ref{eq:strain-numerical}), this condition becomes 
\begin{equation}
	|\epsilon_0| \lesssim M_p \sqrt{\frac{h_{\text{min}} m_\omega R}{8\pi}}. \label{eq:epsilon-bound}
\end{equation}

For a typical LIGO/Virgo event, we consider 
 \(h_{\text{min}} \sim 10^{-23}\) (peak strain sensitivity for GW150914-like 
events), while \(R \sim 400\) Mpc \(\sim 1.2 \times 10^{25}\) m (luminosity 
distance to a typical binary merger). Furthermore, since
 \(m_\omega\) is unknown, we can consider two limiting cases.
 
	  If \(m_\omega\) is very small (ultralight), the bound on \(|\epsilon_0|\) 
becomes tighter because the denominator is smaller. For \(m_\omega \sim 
10^{-14}\) eV \(\sim 1.6 \times 10^{-29}\) kg \(\sim 1.6 \times 10^{-50}\) 
m\(^{-1}\) (in natural units, \(m_\omega\) in eV corresponds to \(m_\omega\) in 
m\(^{-1}\) via \(1 \text{ eV} = 5.06 \times 10^6 \text{ m}^{-1}\)),  we have 
\(m_\omega R = (10^{-14} \text{ eV}) \cdot (6.2 \times 10^{31} \text{ eV}^{-1}) 
= 6.2 \times 10^{17}\). Note that  the product \(m_\omega R\) is 
enormous even for tiny masses. Then Eq.~(\ref{eq:epsilon-bound}) gives
		\begin{equation}
			|\epsilon_0| \lesssim M_p \sqrt{ \frac{10^{-23} \cdot 10^{17}}{8\pi} } \sim M_p \sqrt{10^{-7}} \sim 3 \times 10^{-4} M_p. \label{eq:bound-example}
		\end{equation}
		Thus, for an ultralight Weyl field with \(m_\omega \sim 10^{-14}\) eV, 
the constraint on the fluctuation amplitude is \(|\epsilon_0| \lesssim 10^{-4} 
M_p\). This is not a very restrictive bound, and  fluctuations of this size are 
plausible in many scenarios.
		
 On the other hand, if we instead assume the naturalness argument that 
\(|\epsilon_0| \sim M_p\) (i.e.  the Weyl field fluctuates at the Planck 
scale), then  we obtain a bound on the combination 
\(m_\omega R\):
		\begin{equation}
			m_\omega R \gtrsim \frac{8\pi M_p^2}{M_p^2 h_{\text{min}}} = \frac{8\pi}{h_{\text{min}}} \sim 2.5 \times 10^{24}. \label{eq:mass-bound}
		\end{equation}
		For \(R \sim 10^{25}\) m, this implies \(m_\omega \gtrsim 2.5 \times 10^{24} / 10^{25} \text{ m}^{-1} \sim 0.25 \text{ m}^{-1} \sim 1.3 \times 10^6 \text{ eV}\). So for Planck-scale fluctuations, the Weyl mass must be larger than about \(10^6\) eV to avoid overproducing GW strain. This is a significantly stronger requirement.

However, we emphasize that both estimates depend crucially on the assumed value of \(|\epsilon_0|\), which is not fixed by the theory. In the absence of a first-principles calculation of \(\epsilon_0\) (which would require a full quantum treatment of the Weyl field), we cannot derive a robust bound on \(m_\omega\) or \(\alpha\) alone. The best we can say is that, for a given \(|\epsilon_0|\), current GW observations place an upper bound on the combination \(|\epsilon_0|^2/(m_\omega R)\).

The estimates above lead to the following conclusions, each accompanied by 
important caveats.
  If the Weyl field fluctuations are small (\(|\epsilon_0| \ll M_p\)), the 
induced strain is tiny and easily compatible with current observations. In this 
case, no meaningful constraints on \(m_\omega\) or \(\alpha\) emerge. Moreover, 
if the Weyl field fluctuations are Planckian (\(|\epsilon_0| \sim 
M_p\)), the induced strain would be large unless \(m_\omega\) is sufficiently 
large. This gives \(m_\omega \gtrsim 10^6\) eV. However, a Planckian fluctuation 
of a massive vector field would carry enormous energy density, likely 
back-reacting strongly on the background geometry and invalidating the 
perturbative expansion. Thus, the assumption \(|\epsilon_0| \sim M_p\) may be 
inconsistent with the weak-field approximation used throughout this work.
	
On the other hand, in neither case do we obtain a direct bound on the Weyl 
coupling constant \(\alpha\) alone. The mass \(m_\omega\) is related to 
\(\alpha\) and the symmetry-breaking scale via Eq.~(\ref{eq:mass-relation}) 
\cite{Drechsler:1998gy}, but without knowledge of the symmetry-breaking scale, 
this does not translate into a bound on \(\alpha\).
	Lastly, degeneracies with source parameters: In realistic GW data analysis, 
an additional contribution to the strain is degenerate with the source 
distance, inclination angle, and calibration uncertainties. Therefore, the 
simple comparison \(|\Delta h| < h\).

\section{conclusions}\label{Conclusions}

In this work we have investigated how spacetime non-metricity, realized minimally through vectorial Weyl geometry, can be probed by two of the most powerful tools of modern precision physics: atomic clocks and gravitational-wave (GW) detectors. Our analysis leads to a set of distinct and complementary conclusions.

We first demonstrated that the observable signature of non-metricity is a gauge-invariant, path-dependent change in physical lengths, described by the flux of the Weyl field strength \(F_{\mu\nu}\) through the area enclosed by two trajectories. This formulation cleanly separates the physical effect from the unobservable gauge field, providing a robust framework for experimental tests.

Exploiting the extraordinary sensitivity of modern atomic clocks, which can resolve fractional frequency shifts at the \(10^{-18}\) level, we derived stringent constraints. For ultralight Weyl fields with masses below the critical scale \(m_\omega \lesssim 10^{-13} ~\text{eV}\), terrestrial clock-comparison experiments bound the Weyl gauge coupling to \(\alpha \lesssim 10^{-9}\). Heavier fields evade these constraints through an exponential Yukawa suppression, rendering them insensitive to Earth-scale experiments.

Turning to GWs, we established that, at linear order, the Weyl and gravitational sectors decouple: GWs do not directly source the Weyl gauge field through curvature or connection couplings. However, the dynamical Weyl field, if present, carries a stress-energy tensor that back-reacts on the metric at second order in perturbation theory. This mechanism introduces an anomalous strain, \(\Delta h\), whose magnitude is proportional to \(|\epsilon_0|^2/(m_\omega R)\). By requiring this correction to remain below the detection thresholds of current LIGO/Virgo observations, we obtained initial, order-of-magnitude bounds linked to the fluctuation amplitude of the Weyl field.

Our results collectively demonstrate that the absence of observed deviations from General Relativity in current precision and gravitational-wave experiments should not be misinterpreted as a null result for non-metricity. Instead, these observations already place meaningful constraints on a well-defined class of non-metric degrees of freedom. Furthermore, atomic clocks and gravitational-wave interferometers act in synergy, probing complementary regions of the parameter space in mass and coupling strength. With the anticipated advent of next-generation gravitational-wave observatories and further improvements in clock technologies, the search for the subtle geometric fingerprints of non-metricity is only just beginning.

Finally, several possibilities for future work are opened by the present study.
Although our explicit analysis has focused on Weyl geometry, the underlying
strategy of connecting non-metricity to precision metrology and gravitational
waves is more general.
Extensions to other realizations of non-metricity, including metric-affine and
symmetric teleparallel theories such as $f(Q)$ gravity, are well motivated, but
require a dedicated treatment of matter couplings and of the corresponding
observable map.
In particular, in non-vectorial cases the relation between length transport,
clock measurements, and gravitational-wave propagation is expected to be more
intricate, however having in mind the broad interest for   $f(Q)$ theories and 
their applications  
\cite{Heisenberg:2023lru,Anagnostopoulos:2021ydo,Lazkoz:2019sjl,Lu:2019hra,
	Mandal:2020buf, Frusciante:2021sio, Khodadi:2025upl,
	Gadbail:2022jco, Khyllep:2021pcu,Barros:2020bgg,Shabani:2023xfn, 
	De:2023xua,
	Dimakis:2021gby,Anagnostopoulos:2022gej,Guzman:2024cwa, 
	Boiza:2025xpn,  Basilakos:2025olm}   such an analysis is both necessary and 
interesting.
We leave a systematic exploration of these directions for future work, and
anticipate that the rapidly advancing precision of clocks and gravitational-wave
detectors will continue to provide powerful tools for probing the geometric
structure of spacetime beyond metric compatibility.

\begin{acknowledgments}
The authors would like to thank Tiberiu Harko for insightful discussions on the 
initial draft.
E.N.S. gratefully acknowledges  the 
contribution of 
the LISA Cosmology Working Group (CosWG), as well as support from the COST 
Actions CA21136 -  Addressing observational tensions in cosmology with 
systematics and fundamental physics (CosmoVerse)  - CA23130, Bridging 
high and low energies in search of quantum gravity (BridgeQG)  and CA21106 -  
 COSMIC WISPers in the Dark Universe: Theory, astrophysics and 
experiments (CosmicWISPers). 
\end{acknowledgments}

\appendix 

\appendix

\section{Direct curvature sourcing of the Weyl field}
\label{app:A}

A natural mechanism through which GWs could excite the Weyl gauge field \(\omega_\mu\) is via curvature itself. In this Appendix we examine this possibility explicitly and show that, although curvature does source \(\omega_\mu\) in Weyl quadratic gravity, such a mechanism does not operate for linearized GWs in vacuum. We present a detailed, step-by-step derivation to make the calculation transparent and to identify precisely where the coupling vanishes.

We begin with the Weyl connection, which is given by
\begin{equation}
	\tilde{\Gamma}_{\mu\nu}^\lambda = \Gamma_{\mu\nu}^\lambda + \frac{\alpha}{2}\left(\delta_\mu^\lambda \omega_\nu + \delta_\nu^\lambda \omega_\mu - g_{\mu\nu}\omega^\lambda\right), \label{eq:weyl-connection}
\end{equation}
where \(\Gamma_{\mu\nu}^\lambda\) is the Levi-Civita connection of the metric \(g_{\mu\nu}\). The Weyl scalar curvature \(\tilde{R}\) is obtained by contracting the Weyl Ricci tensor \(\tilde{R}_{\mu\nu} = \tilde{R}^\rho{}_{\mu\rho\nu}\), where \(\tilde{R}^\rho{}_{\sigma\mu\nu}\) is the curvature tensor of the Weyl connection.

The relation between \(\tilde{R}\) and the Riemannian Ricci scalar \(R\) is known in the literature~\cite{Ghilencea:2018dqd,Ghilencea:2022lcl}:
\begin{equation}
	\tilde{R} = R - 3\alpha \nabla_\mu \omega^\mu - \frac{3}{2}\alpha^2 \omega_\mu \omega^\mu. \label{eq:R-tilde-relation}
\end{equation}
We will verify this relation to linear order to ensure no sign errors. Starting from the Weyl connection (\ref{eq:weyl-connection}), the difference between the Weyl and Levi-Civita connections is the tensor
\begin{equation}
	\Delta_{\mu\nu}^\lambda = \tilde{\Gamma}_{\mu\nu}^\lambda - \Gamma_{\mu\nu}^\lambda = \frac{\alpha}{2}\left(\delta_\mu^\lambda \omega_\nu + \delta_\nu^\lambda \omega_\mu - g_{\mu\nu}\omega^\lambda\right). \label{eq:Delta}
\end{equation}
The Weyl curvature tensor can be expressed in terms of the Riemannian curvature tensor \(R^\rho{}_{\sigma\mu\nu}\) and covariant derivatives of \(\Delta\):
\begin{equation}
	\tilde{R}^\rho{}_{\sigma\mu\nu} = R^\rho{}_{\sigma\mu\nu} + \nabla_\mu \Delta^\rho_{\nu\sigma} - \nabla_\nu \Delta^\rho_{\mu\sigma} + \Delta^\rho_{\mu\lambda}\Delta^\lambda_{\nu\sigma} - \Delta^\rho_{\nu\lambda}\Delta^\lambda_{\mu\sigma}. \label{eq:curvature-decomposition}
\end{equation}
Contracting to obtain the Ricci tensor \(\tilde{R}_{\mu\nu} = \tilde{R}^\rho{}_{\mu\rho\nu}\):
\begin{equation}
	\tilde{R}_{\mu\nu} = R_{\mu\nu} + \nabla_\rho \Delta^\rho_{\nu\mu} - \nabla_\nu \Delta^\rho_{\rho\mu} + \Delta^\rho_{\rho\lambda}\Delta^\lambda_{\nu\mu} - \Delta^\rho_{\nu\lambda}\Delta^\lambda_{\rho\mu}. \label{eq:ricci-decomposition}
\end{equation}
Now we compute the necessary contractions of \(\Delta\). Using (\ref{eq:Delta}):
\begin{align}
	\Delta^\rho_{\nu\mu} &= \frac{\alpha}{2}\left(\delta^\rho_\nu \omega_\mu + \delta^\rho_\mu \omega_\nu - g_{\nu\mu}\omega^\rho\right), \label{eq:Delta-contraction1} \\
	\Delta^\rho_{\rho\mu} &= 2\alpha \omega_\mu, \label{eq:Delta-contraction2} \\
	\Delta^\rho_{\rho\lambda} &= 2\alpha \omega_\lambda. \label{eq:Delta-contraction3}
\end{align}
Also note that \(\nabla_\mu \omega_\nu\) is the covariant derivative with respect to the Levi-Civita connection.

Substituting into (\ref{eq:ricci-decomposition}) and keeping terms up to linear order in \(\omega_\mu\) (dropping quadratic terms \(\Delta^2\)):
\begin{align}
	\tilde{R}_{\mu\nu} = R_{\mu\nu} + \nabla_\rho\left[\frac{\alpha}{2}\left(\delta^\rho_\nu \omega_\mu + \delta^\rho_\mu \omega_\nu - g_{\nu\mu}\omega^\rho\right)\right] - \nonumber \\ \nabla_\nu(2\alpha \omega_\mu) + 
	\mathcal{O}(\omega^2), \label{eq:ricci-linear}
\end{align}
and expanding the covariant derivatives we find  
\begin{equation}
	\tilde{R}_{\mu\nu}  	 = R_{\mu\nu} - \frac{3\alpha}{2}\nabla_\nu 
\omega_\mu + \frac{\alpha}{2}\nabla_\mu \omega_\nu - 
\frac{\alpha}{2}g_{\nu\mu}\nabla_\rho \omega^\rho + \mathcal{O}(\omega^2). 
\label{eq:ricci-linear-final}
\end{equation}
Additionally, contracting with \(g^{\mu\nu}\) to obtain the Weyl scalar 
curvature \(\tilde{R} = g^{\mu\nu}\tilde{R}_{\mu\nu}\) we find 
\begin{align}
	\tilde{R} = g^{\mu\nu}\tilde{R}_{\mu\nu} = (\eta^{\mu\nu} - h^{\mu\nu})(R_{\mu\nu} + \delta R_{\mu\nu}^{(\omega)} + \cdots) = \nonumber \\
	 \eta^{\mu\nu}R_{\mu\nu} + \eta^{\mu\nu}\delta R_{\mu\nu}^{(\omega)} + 
\mathcal{O}(h\omega, h^2, \omega^2) , \label{eq:scalar-contraction}
\end{align}
where  
\begin{align}
	\eta^{\mu\nu}\delta R_{\mu\nu}^{(\omega)} &= 
\eta^{\mu\nu}\left(-\frac{3\alpha}{2}\nabla_\nu \omega_\mu + 
\frac{\alpha}{2}\nabla_\mu \omega_\nu - \frac{\alpha}{2}\eta_{\nu\mu}\nabla_\rho 
\omega^\rho\right) \nonumber \\ 	&= -3\alpha \nabla_\mu\omega^\mu. 
\label{eq:contraction-result}
\end{align}

Now, the Weyl quadratic gravity action contains the term
\begin{equation}
	\mathcal{L}_{\tilde{R}^2} = \frac{1}{4!\xi^2} \sqrt{-g} \tilde{R}^2. \label{eq:R2-lagrangian}
\end{equation}
To derive the equation of motion for \(\omega_\mu\), we first express 
\(\tilde{R}^2\) in terms of Riemannian quantities. Using 
(\ref{eq:R-tilde-relation}) and keeping terms up to quadratic order in 
perturbations: 
\begin{align}
	\tilde{R}^2 &= R^2 - 6\alpha R \nabla_\mu \omega^\mu - 3\alpha^2 R \omega_\mu \omega^\mu \nonumber \\
	&\quad + 9\alpha^2 (\nabla_\mu \omega^\mu)^2 + 9\alpha^3 (\nabla_\mu \omega^\mu)(\omega_\nu \omega^\nu) + \frac{9}{4}\alpha^4 (\omega_\mu \omega^\mu)^2. \label{eq:R2-full-expansion}
\end{align}
For vacuum gravitational waves, \(R = 0\). In the linearized regime (keeping only terms quadratic in perturbations, since the action is quadratic in fields), the relevant term is
\begin{equation}
	\tilde{R}^2 \supset 9\alpha^2 (\partial_\mu \epsilon^\mu)^2, \label{eq:R2-vacuum}
\end{equation}
where we have replaced \(\nabla_\mu\) with \(\partial_\mu\) because the difference is of higher order in perturbations (the Christoffel symbols are of order \(h\)). 

Now  in order to vary the action with respect to \(\omega_\mu\)  we will use 
the relation
\begin{equation}
	\frac{\delta}{\delta \omega_\nu} \int d^4x \frac{9\alpha^2}{4!\xi^2} 
(\partial_\mu \omega^\mu)^2 = -\frac{3\alpha^2}{4\xi^2} \partial^\nu 
(\partial_\mu \omega^\mu).\label{eq:R2-eom}
\end{equation}

The full action for the Weyl field (including the kinetic term and the \(\tilde{R}^2\) term) is
\begin{equation}
	\mathcal{L}_\omega = \sqrt{-g}\left(-\frac{1}{4} F_{\mu\nu}F^{\mu\nu} + \frac{1}{4!\xi^2}\tilde{R}^2\right). \label{eq:full-omega-lagrangian}
\end{equation}
The variation of the Maxwell term \(-\frac{1}{4}F_{\mu\nu}F^{\mu\nu}\) with respect to \(\omega_\mu\) gives the standard Maxwell equation (in flat spacetime, with \(\sqrt{-g} \approx 1\)):
\begin{equation}
	\frac{\delta}{\delta \omega_\nu} \int d^4x \left(-\frac{1}{4}F_{\mu\nu}F^{\mu\nu}\right) = \partial_\mu F^{\mu\nu}. \label{eq:maxwell-variation}
\end{equation}
Combining with the \(\tilde{R}^2\) contribution, the linearized equation of motion (to first order in \(\epsilon_\mu\)) is
\begin{equation}
	\partial_\mu F^{\mu\nu} - \frac{3\alpha^2}{4\xi^2} \partial^\nu 
(\partial_\mu \epsilon^\mu) = 0, \label{eq:linear-eom}
\end{equation}
and recalling that \(\partial_\mu F^{\mu\nu} = \square \epsilon^\nu - 
\partial^\nu (\partial_\mu \epsilon^\mu)\) we find 
\begin{equation}
	\left[1 - \left(1 + \frac{3\alpha^2}{4\xi^2}\right)\right] \square (\partial_\mu \epsilon^\mu) = -\frac{3\alpha^2}{4\xi^2} \square (\partial_\mu \epsilon^\mu) = 0. \label{eq:divergence-result}
\end{equation}
Thus, for \(\alpha \neq 0\), we obtain \(\square (\partial_\mu \epsilon^\mu) = 0\). This implies that \(\partial_\mu \epsilon^\mu\) is a free massless scalar field. In the absence of sources, we can consistently set \(\partial_\mu \epsilon^\mu = 0\) (this is a choice of gauge, analogous to the Lorenz gauge for the Proca field). With this choice, the equation of motion reduces to
\begin{equation}
	\square \epsilon^\nu = 0. \label{eq:wave-eq}
\end{equation}
Thus, the Weyl field satisfies the massless wave equation. The \(\tilde{R}^2\) term modifies the longitudinal mode dynamics but does not introduce a source term coupling to \(h_{\mu\nu}\). Importantly, there is no term proportional to the Riemannian curvature \(R\) or its derivatives that would source \(\epsilon_\mu\) linearly, because \(R\) vanishes for vacuum GWs.

We close this Appendix by mentioning that the action contains only 
\(\tilde{R}^2\) and 
\(\tilde{C}^2\), but no  \(R\tilde{R}\) 
term.  The \(\tilde{C}^2\) term does not produce a linear coupling 
either, as shown in Appendix~\ref{app:B}. Even if we considered a more general 
action, any coupling of the form \(R \tilde{R}\) would involve the Riemannian 
curvature \(R\), which is zero for vacuum GWs at linear order. At second order 
in perturbations, \(R \sim \mathcal{O}(h^2)\), and the corresponding source term 
would be of order \(h^2\). For detectable GWs with \(h \sim 10^{-21}\), \(h^2 
\sim 10^{-42}\), which is completely negligible. Therefore, direct curvature 
sourcing does not provide an efficient mechanism for exciting the Weyl field in 
the context of gravitational-wave propagation.

\section{Direct coupling through the Weyl connection and curvature invariants}
\label{app:B}

In this Appendix, we examine whether GWs can excite the Weyl gauge field \(\omega_\mu\) through direct couplings arising from the Weyl connection and the associated curvature tensors beyond the scalar curvature. Specifically, we analyze the \(\tilde{C}^2\) term and other possible quadratic invariants. The purpose is to verify explicitly that, even when such geometric couplings are present in the action, no direct sourcing of \(\omega_\mu\) by GWs occurs at linear order.

We expand around Minkowski spacetime with a vanishing background Weyl field:
\begin{equation}
	g_{\mu\nu} = \eta_{\mu\nu} + h_{\mu\nu}, \qquad \omega_\mu = \epsilon_\mu, \qquad |h_{\mu\nu}| \ll 1,\ |\epsilon_\mu| \ll 1. \label{eq:expansion-B}
\end{equation}
The Levi-Civita connection to first order in \(h_{\mu\nu}\) is
\begin{equation}
	\Gamma_{\mu\nu}^\lambda = \frac{1}{2}\eta^{\lambda\rho}(\partial_\mu h_{\nu\rho} + \partial_\nu h_{\mu\rho} - \partial_\rho h_{\mu\nu}) + \mathcal{O}(h^2). \label{eq:christoffel-linear}
\end{equation}
The Weyl connection from Eq.~(\ref{eq:weyl-connection}) becomes
\begin{align}
	\tilde{\Gamma}_{\mu\nu}^\lambda &= \Gamma_{\mu\nu}^\lambda + \frac{\alpha}{2}\left(\delta_\mu^\lambda \epsilon_\nu + \delta_\nu^\lambda \epsilon_\mu - \eta_{\mu\nu}\epsilon^\lambda\right) \nonumber \\
	&\quad - \frac{\alpha}{2}\left(h_{\mu\nu}\epsilon^\lambda + \eta_{\mu\nu}h^{\lambda\rho}\epsilon_\rho\right) + \mathcal{O}(h^2, \epsilon^2, h\epsilon), \label{eq:weyl-connection-linear}
\end{align}
where the last term comes from expanding \(g_{\mu\nu}\omega^\lambda = (\eta_{\mu\nu}+h_{\mu\nu})(\epsilon^\lambda - h^{\lambda\rho}\epsilon_\rho + \cdots)\) to first order in perturbations. For the purpose of identifying linear couplings, we only need the terms that are linear in \(h\) and \(\epsilon\) separately. The term \(\frac{\alpha}{2}(- \eta_{\mu\nu}h^{\lambda\rho}\epsilon_\rho)\) is of order \(h\epsilon\) and thus bilinear; it will not contribute to linear sourcing.

Now, the Weyl curvature tensor is given by
\begin{equation}
	\tilde{R}^\rho{}_{\sigma\mu\nu} = \partial_\mu \tilde{\Gamma}_{\nu\sigma}^\rho - \partial_\nu \tilde{\Gamma}_{\mu\sigma}^\rho + \tilde{\Gamma}_{\mu\lambda}^\rho \tilde{\Gamma}_{\nu\sigma}^\lambda - \tilde{\Gamma}_{\nu\lambda}^\rho \tilde{\Gamma}_{\mu\sigma}^\lambda. \label{eq:weyl-curvature-def}
\end{equation}
Keeping only terms linear in \(h\) and \(\epsilon\) (i.e., ignoring products of perturbations), the quadratic terms \(\tilde{\Gamma}\tilde{\Gamma}\) are of order \(h^2\), \(\epsilon^2\), or \(h\epsilon\), and thus do not contribute to the linearized equations. Therefore,
\begin{equation}
	\tilde{R}^\rho{}_{\sigma\mu\nu} = \partial_\mu \tilde{\Gamma}_{\nu\sigma}^\rho - \partial_\nu \tilde{\Gamma}_{\mu\sigma}^\rho + \mathcal{O}(\text{quadratic}). \label{eq:curvature-linear}
\end{equation}
Substituting the linearized connection from Eq.~(\ref{eq:weyl-connection-linear}):
\begin{align}
	\tilde{R}^\rho{}_{\sigma\mu\nu} &= \partial_\mu \Gamma_{\nu\sigma}^\rho - \partial_\nu \Gamma_{\mu\sigma}^\rho \nonumber \\
	&\quad + \frac{\alpha}{2}\partial_\mu\left(\delta_\nu^\rho \epsilon_\sigma + \delta_\sigma^\rho \epsilon_\nu - \eta_{\nu\sigma}\epsilon^\rho\right) \nonumber \\
	&\quad - \frac{\alpha}{2}\partial_\nu\left(\delta_\mu^\rho \epsilon_\sigma + \delta_\sigma^\rho \epsilon_\mu - \eta_{\mu\sigma}\epsilon^\rho\right) + \mathcal{O}(\text{quadratic}). \label{eq:curvature-expanded}
\end{align}
The first line is precisely the Riemannian curvature tensor \(R^\rho{}_{\sigma\mu\nu}\) (linearized). The remaining terms are the contributions from the Weyl field.

Now contracting to obtain the Ricci tensor  \(\tilde{R}_{\mu\nu} = 
\tilde{R}^\rho{}_{\mu\rho\nu}\) 
we find  
\begin{align}
	\tilde{R}_{\mu\nu} &= R_{\mu\nu} + \frac{\alpha}{2}\left(\partial_\mu \epsilon_\nu + \partial_\nu \epsilon_\mu - \eta_{\mu\nu}\partial_\rho\epsilon^\rho\right) \nonumber \\
	&\quad - \frac{\alpha}{2}\left(\partial_\nu \epsilon_\mu + 4\partial_\nu 
\epsilon_\mu - \partial_\nu \epsilon_\mu\right) + \mathcal{O}(\text{quadratic}) 
\nonumber \\  
	&= R_{\mu\nu} + \frac{\alpha}{2}\partial_\mu \epsilon_\nu - \frac{3\alpha}{2}\partial_\nu \epsilon_\mu - \frac{\alpha}{2}\eta_{\mu\nu}\partial_\rho\epsilon^\rho. \label{eq:ricci-final-linear}
\end{align}
This matches our earlier calculation in Appendix~\ref{app:A} (with \(\nabla\) 
replaced by \(\partial\) at linear order).  Finally, contracting with 
\(\eta^{\mu\nu}\) to obtain the scalar curvature:
\begin{equation}
	\tilde{R}  =   R - 3\alpha \partial_\mu\epsilon^\mu. 
\label{eq:scalar-final}
\end{equation}

 Proceeding forward, the Weyl tensor squared term is an important part of the 
action. Thus, we need to examine whether it can produce a linear coupling 
between \(h_{\mu\nu}\) and \(\epsilon_\mu\). The Weyl tensor 
\(\tilde{C}_{\mu\nu\rho\sigma}\) is defined as the traceless part of the 
curvature tensor. In Weyl geometry, it satisfies the 
identity~\cite{Drechsler:1998gy}
\begin{equation}
	\tilde{C}_{\mu\nu\rho\sigma}\tilde{C}^{\mu\nu\rho\sigma} = C_{\mu\nu\rho\sigma}C^{\mu\nu\rho\sigma} + \frac{3}{2}\alpha^2 F_{\mu\nu}F^{\mu\nu} + \mathcal{O}(\alpha^3), \label{eq:C2-identity}
\end{equation}
where \(C_{\mu\nu\rho\sigma}\) is the Riemannian Weyl tensor (constructed from the Levi-Civita connection). This identity is crucial: it shows that the \(\tilde{C}^2\) term decomposes into a purely Riemannian part (which depends only on \(h_{\mu\nu}\) and gives fourth-order equations for the metric) and a Maxwell term for the Weyl field (which depends only on \(\epsilon_\mu\)). There is no cross term linear in both \(h\) and \(\epsilon\). The \(\mathcal{O}(\alpha^3)\) terms are of higher order in the coupling and also do not produce linear couplings.

We can verify this identity to first order in \(\alpha\) by explicit expansion. The Weyl tensor can be written as
\begin{eqnarray}
&&\!\!\!\! 	\tilde{C}_{\mu\nu\rho\sigma} = C_{\mu\nu\rho\sigma} + 
\frac{\alpha}{2}\left( \eta_{\mu\rho}F_{\nu\sigma} + \eta_{\nu\sigma}F_{\mu\rho} 
- \eta_{\mu\sigma}F_{\nu\rho} - \eta_{\nu\rho}F_{\mu\sigma} \right)
\nonumber\\
&& \ \ \ \ \ \ \ \ \   + 
\mathcal{O}(\alpha^2), \label{eq:C-expansion}
\end{eqnarray}
where \(F_{\mu\nu} = \partial_\mu\epsilon_\nu - \partial_\nu\epsilon_\mu\) is the field strength. Squaring this expression and keeping terms up to \(\mathcal{O}(\alpha^2)\):
\begin{eqnarray}
	&& \!\!\!\! \! \! \!    \tilde{C}^2 = C^2 + \alpha \left[ 
C^{\mu\nu\rho\sigma}\left( \eta_{\mu\rho}F_{\nu\sigma} + 
\eta_{\nu\sigma}F_{\mu\rho}- 
\eta_{\mu\sigma}F_{\nu\rho}
\right.\right.
\nonumber\\
&& \ \ \ \ \ \ \ \ \  \ \ \   \ \  \left.\left.  - 
\eta_{\nu\rho}F_{\mu\sigma} \right) \right]   + \frac{\alpha^2}{4}\left( 
\eta_{\mu\rho}F_{\nu\sigma} + \cdots \right)^2. \label{eq:C2-expansion}
\end{eqnarray}
The linear term in \(\alpha\) involves the contraction of the Riemannian Weyl 
tensor  \(C_{\mu\nu\rho\sigma}\) with \(F_{\mu\nu}\). However, the Riemannian 
Weyl tensor   is traceless and satisfies the Bianchi 
identities. In vacuum, \(C_{\mu\nu\rho\sigma}\) is nonzero for gravitational 
waves, but the specific contraction with \(F_{\mu\nu}\)   vanishes (since
	$C^{\mu\nu\rho\sigma}\eta_{\mu\rho}F_{\nu\sigma} = C^{\mu\nu\rho}{}_\mu 
F_{\nu\rho} = C^{\nu\rho}{}F_{\nu\rho}=0$). Similarly, all other contractions in 
the linear term 
involve traces of the Weyl tensor and thus vanish. Hence, the linear term is 
zero. The next term is quadratic in \(F_{\mu\nu}\) and gives the 
\(\frac{3}{2}\alpha^2 F_{\mu\nu}F^{\mu\nu}\) term after evaluating the 
combinatorial factors. This confirms the identity.

In summary, from the detailed calculations above (Appendices \ref{app:A} and 
\ref{app:B}), we result to the following. 
 The Weyl scalar curvature \(\tilde{R}\) contains a term linear in 
\(\partial_\mu\epsilon^\mu\) but no term linear in \(h_{\mu\nu}\) for vacuum GWs 
(since \(R=0\)). Moreover, 
the \(\tilde{R}^2\) term in the action gives rise to a contribution to the Weyl 
field equation that modifies the longitudinal mode dynamics but does not 
introduce a source term proportional to \(h_{\mu\nu}\).
  The \(\tilde{C}^2\) term decomposes into a Riemannian \(C^2\) term (which 
governs the dynamics of \(h_{\mu\nu}\) at fourth order) and a Maxwell term for 
\(\epsilon_\mu\), with no linear cross-coupling. Additionally,  
any other quadratic curvature invariant (e.g., 
\(\tilde{R}_{\mu\nu}\tilde{R}^{\mu\nu}\)) can be expressed as linear 
combinations of \(\tilde{R}^2\) and \(\tilde{C}^2\) in four dimensions (up to 
topological terms), so no new couplings appear.
 Therefore, GWs do not directly source the Weyl gauge field at linear order 
through curvature or connection couplings. This negative result is robust and 
closes several seemingly plausible coupling channels.
 
 Hence, this analysis justifies our focus in Section~\ref{GWprobes} on the 
backreaction mechanism, where the Weyl field is treated as an independent 
dynamical degree of freedom whose stress-energy tensor sources GWs at second 
order.

\bibliographystyle{apalike}

\end{document}